\RequirePackage{fix-cm}
\RequirePackage{amsmath}
\documentclass{svjour3}                     
\smartqed  
\usepackage{amssymb}
\usepackage{amsmath}
\usepackage{color}

\def\L{{\mathcal L}}
\newtheorem{thm}{Proposition}

\usepackage[justification=centering]{caption}
\usepackage[caption=false,font=footnotesize]{subfig}
\usepackage{graphicx}
\begin{document}

\title{Co-Prime Sensing with Multiple Periods and Difference Set Analysis in the Presence of Sampling Jitter}



\author{Usham V. Dias}


\institute{Usham V. Dias \at
              Dept. Electrical Engineering,  \\
              Indian Institute of Technology Delhi\\
              New Delhi, India\\
              \email{ushamdias@gmail.com}           
}


\maketitle

\begin{abstract}
Co-prime arrays and samplers with multiple periods is a framework in which the co-prime structure is repeated multiple times. In this paper, the effects of perturbations in sampling locations on the difference set of the prototype co-prime structure with multiple periods is analysed. Based on this analysis, a method to estimate the autocorrelation that maximizes the amount of information extracted from the data is proposed. The advantage is limited only to samplers, and is not observed in antenna arrays. The expression for the number of additional contributors available for estimation is derived. The contributors increase with the increase in the number of periods. In addition, the expressions for the computational complexity are derived in the presence of jitter. This provides an upper bound on the number of multiplications and additions for hardware implementation. 
\end{abstract}

\section{Introduction}
\label{intro}
Analog-to-Digital Conversion (or the sampling process) may encounter errors in the amplitude of the acquired signal and/or error in timing. The work in \cite{Jjitter} focuses on timing jitter errors. It assumes a random jitter model and studies its properties for stochastic and deterministic signals. Jitter may selectively attenuate the spectral distribution and is undesirable. It also discusses optimal linear operation and shows that jitter may not change the nature of the optimal operations.

The work in \cite{13.5} considers independent jitters and procedures for spectral estimation. It studies the autocovariance estimation procedure, properties, and relative efficiency of the estimators.
Timing jitter effects on spatio-frequency covariance matrix with direction and delay information is studied in \cite{13.4}. Jitter variance estimation and compensation methods are discussed.
The work \cite{13.3} studies discrete time observation-based covariance estimation with jitter and delay. It also considers the estimator normality and consistency.
In \cite{13.8}, sampling jitter noise for system identification is considered and mitigation methods are discussed.
Timing jitters are also studied in the field of communication and clock recovery circuits. The work in \cite{Jitters_Naveen} considers data dependent, random, and a combination of the two jitters with a focus on reducing the settling time.

For the case of sub-Nyquist sampled signals, the work in \cite{13.13}, presents a system model for jitter reconstruction. An annihilating filter is used to estimate the jitter timings. The reconstruction of signal is achieved using a Slepian function. With the sampling rate and sub-band information, it provides improvement in the SNR, i.e. Signal-to-Noise ratio.

Sub-Nyquist co-prime and nested structures were studied under perturbed conditions in \cite{4.31,13.1}. It considers both spatial and temporal domains with additive perturbation and sampling time jitter. It shows that the errors in autocorrelation estimation (under certain assumptions) is bounded. Most of the work considers the study of jitters in the statistical sense, however, the work in \cite{U_S_2,UVD_phdthesis} studies the effect of jitter on the difference set of the sub-Nyquist co-prime array. Here, an increase in the number of contributors for estimation is possible only for the case of temporal sampling and fails for antenna arrays. The reasoning behind it is that the array has only one antenna placed at the zeroth location (first antenna). The co-prime sampler has two independent sub-samplers and hence two samples at the zeroth location (with jitters). 

Co-prime structures with multiple periods were considered in \cite{CAMPs,UVD_phdthesis}. It repeats the co-prime structure multiple times.
This paper is dedicated to the study of the difference set of the sub-Nyquist co-prime sampler with multiple periods under the influence of jitter. It may be noted that \cite{U_S_2} is a special case of the work described here, i.e. it has only one period. Summary of the work considered in subsequent sections is given below:
\begin{enumerate}
	\item The effect of sampling jitters on the difference set of the prototype co-prime samplers with multiple periods is analyzed.
	\item The number of distinct values in each set under the influence of jitter is described in Proposition~\ref{prop_Multi_jitter_dofs} (Section~\ref{JITTER_multi_difference_set}).
	\item The number of contributor (weights) for autocorrelation estimation is studied under the influence of jitter. Proposition~\ref{prop_Multi_jitter_wts} gives the weights for the unmapped location and Proposition~\ref{prop_Multi_jitter_wts_mapped} gives the weights for the mapped locations (Section~\ref{JITTER_multi_weight_fn}).
	\item The computational complexity for autocorrelation estimation in the presence of jitter is derived in Section~\ref{JITTER_multi_complexity}.
\end{enumerate}
\begin{figure*}[!t]
	\centering
	\includegraphics[width=0.9\textwidth]{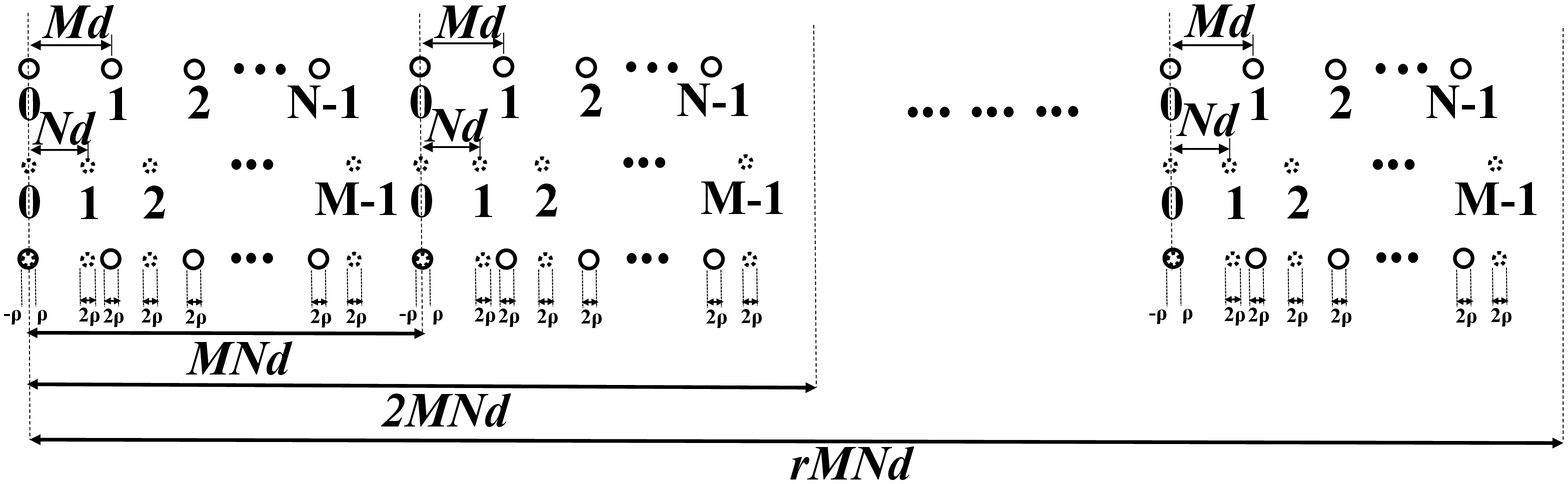}%
	\caption{Co-prime sampling structure with multiple periods in the presence of jitter.}
	\label{jitter_multi_concept}
\end{figure*}
\section{Co-Prime Sampling with Multiple Periods}
\label{JITTER_multi}
A sub-Nyquist co-prime sampler/array is a scheme which has two sub-arrays with inter-element spacing of $Md$ and $Nd$~\cite{4.7}. $d$ is the Nyquist sampling period (or distance), and pair ($M$, $N$) is selected such that they are co-prime. This scheme has many missing values in the acquired data, but the autocorrelation of the sub-Nyquist data can estimate most of the lag values (also refer~\cite{U_S_1,UVD_phdthesis} for low latency coprime-based estimation theory).
The prototype co-prime sampler has been studied in the past. Here, the prototype co-prime sampler with multiple periods is analysed in the presence of jitters. This scheme combines the samples from $r$ co-prime periods to form one snapshot of the acquired signal. It does not disturb the uniform sampling structure of the individual samplers. 
\subsection{Structure under perturbed Condition}
\label{JITTER_multi_structure}
The co-prime structure with jitters is shown in Fig.~\ref{jitter_multi_concept} for multiple periods. Note, the structure with one period, i.e. $MNd$, is the prototype co-prime structure. The actual positions of the sampling times may not be ideal. There can be jitters in the location. This jitter can cause the location to shift by $\rho$ on either side of the true location. $\rho =0$ is an ideal scenario with no jitters. In general, it is assumed that $0 < \rho < \frac{d}{4}$, i.e. $[-\rho, \rho]=[-\frac{d}{4}, \frac{d}{4}]$. The normalized range is $[-\frac{1}{4}, \frac{1}{4}]$. The Nyquist period $d$ can be ignored in general without affecting the discussion. Therefore, the normalized instantaneous jitters $\epsilon_1(n)$ and $\epsilon_2(m)$ produce difference values within the range of $\pm\frac{1}{2}$ about the ideal value. For period $r=1$, the co-prime sampler has one period and is same as the prototype co-prime structure. With $r>1$ the sampler has both of its sub-arrays (or sub-samplers) extended $r$ times. Therefore, the sampling times are given by $Mn$ and $Nm$, where $0\leq n \leq rN-1$ and $0\leq m \leq rM-1$. This means that the signal is captured for $r$ periods, i.e. $rMNd$ seconds, which forms one snapshot for the autocorrelation estimation. The samples acquired by the two independent co-prime samplers coincide at instants $\{0, MN, 2MN, ..., (r-1)MN\}$ for the ideal case, but does not in a jitter perturbed scenario.
\subsection{Difference Set under the influence of Jitter}
\label{JITTER_multi_difference_set}
\begin{figure*}[!t]
	\centering
	\subfloat[${\L}^+_{SM_r}\cup{\L}^-_{SM_r}$]{\includegraphics[width=0.88\textwidth]{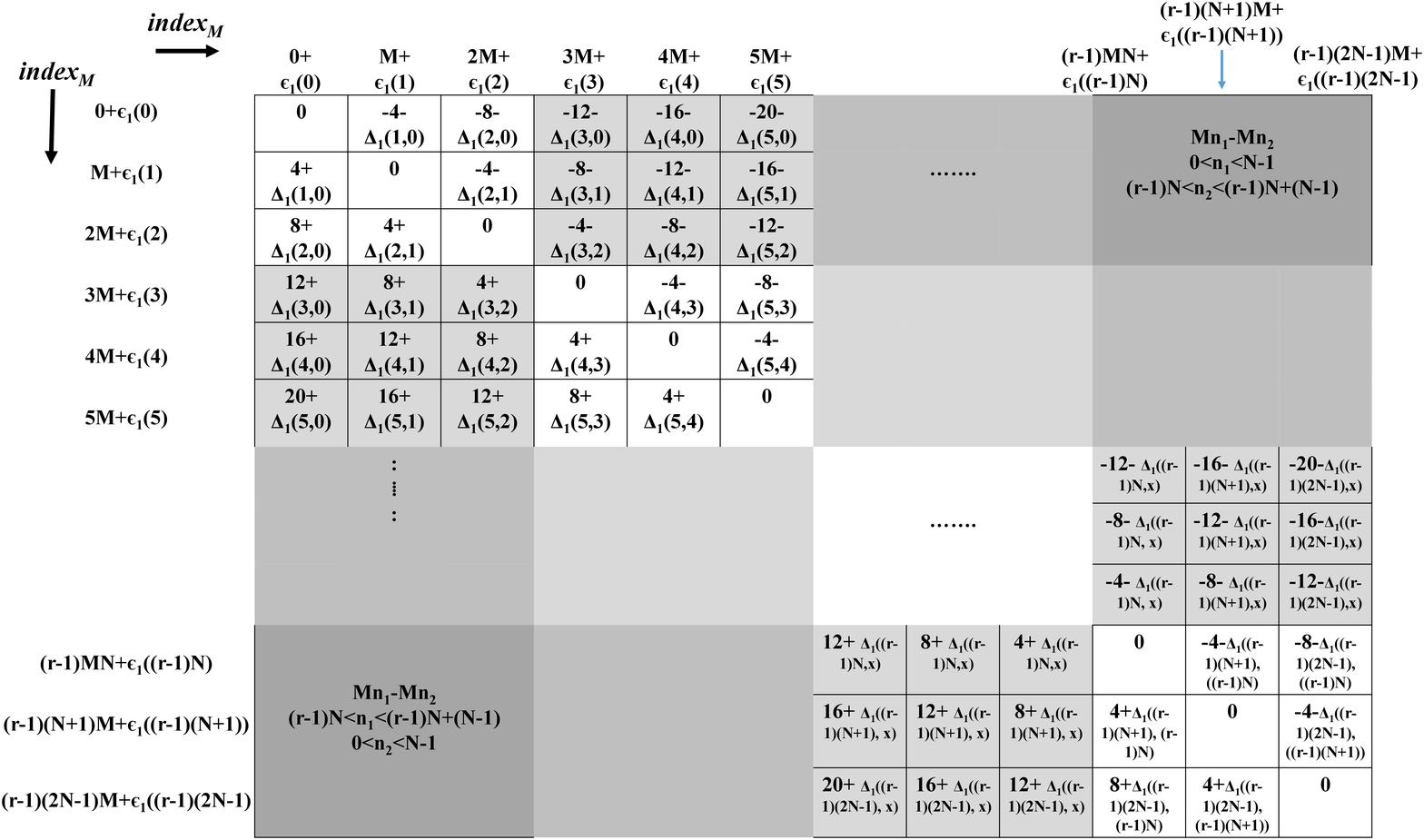}%
		\label{jitter_Multi_LSM}}
	\hfil
	\subfloat[${\L}^+_{SN_r}\cup{\L}^-_{SN_r}$]{\includegraphics[width=0.88\textwidth]{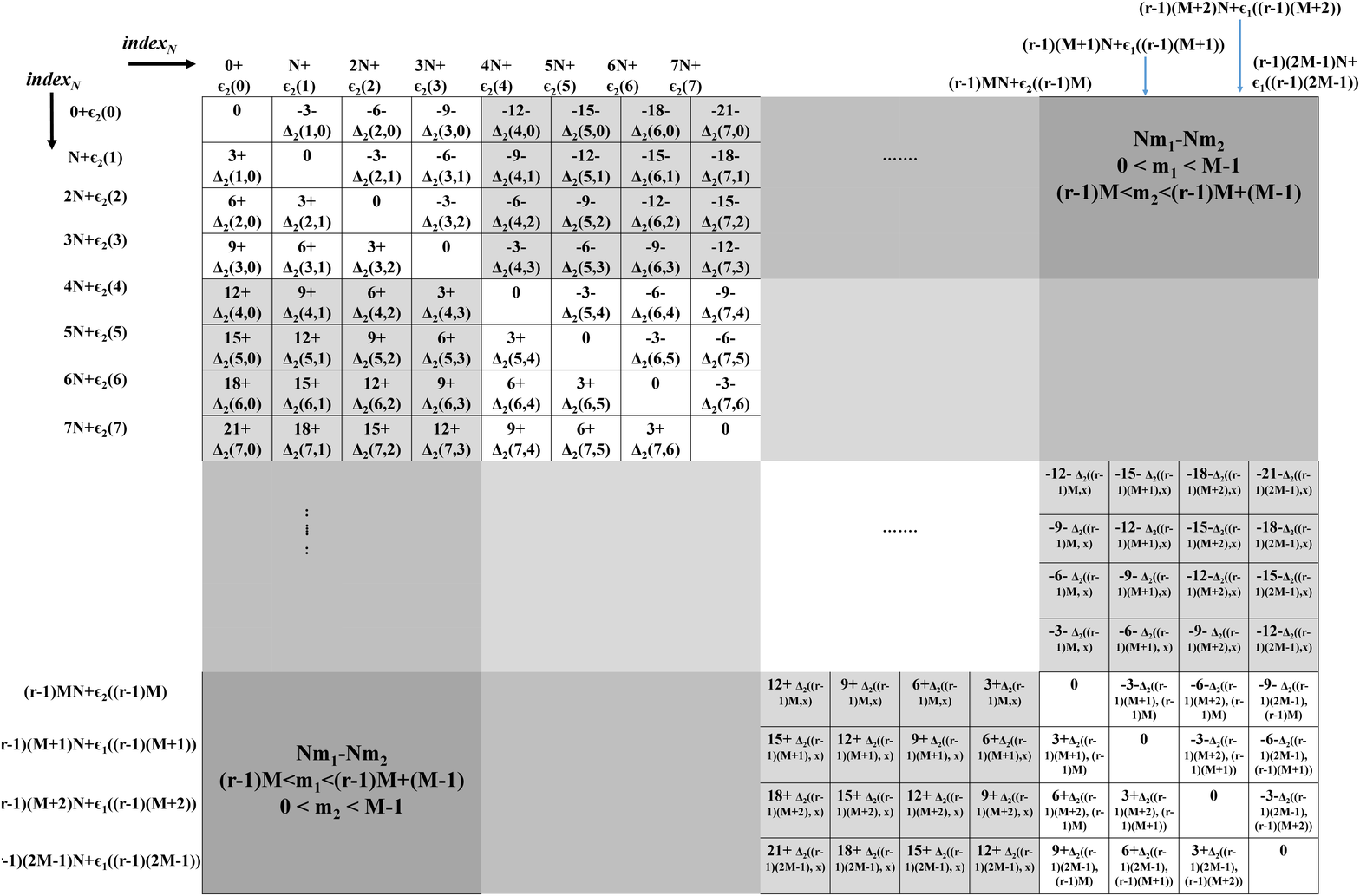}%
		\label{jitter_Multi_LSN}}
	\caption{Co-prime sampling with multiple periods: Self differences in the presence of jitter.}
	\label{jitter_Multi_self_differences}
\end{figure*}
\begin{figure*}[!t]
	\centering
	\includegraphics[width=0.9\textwidth]{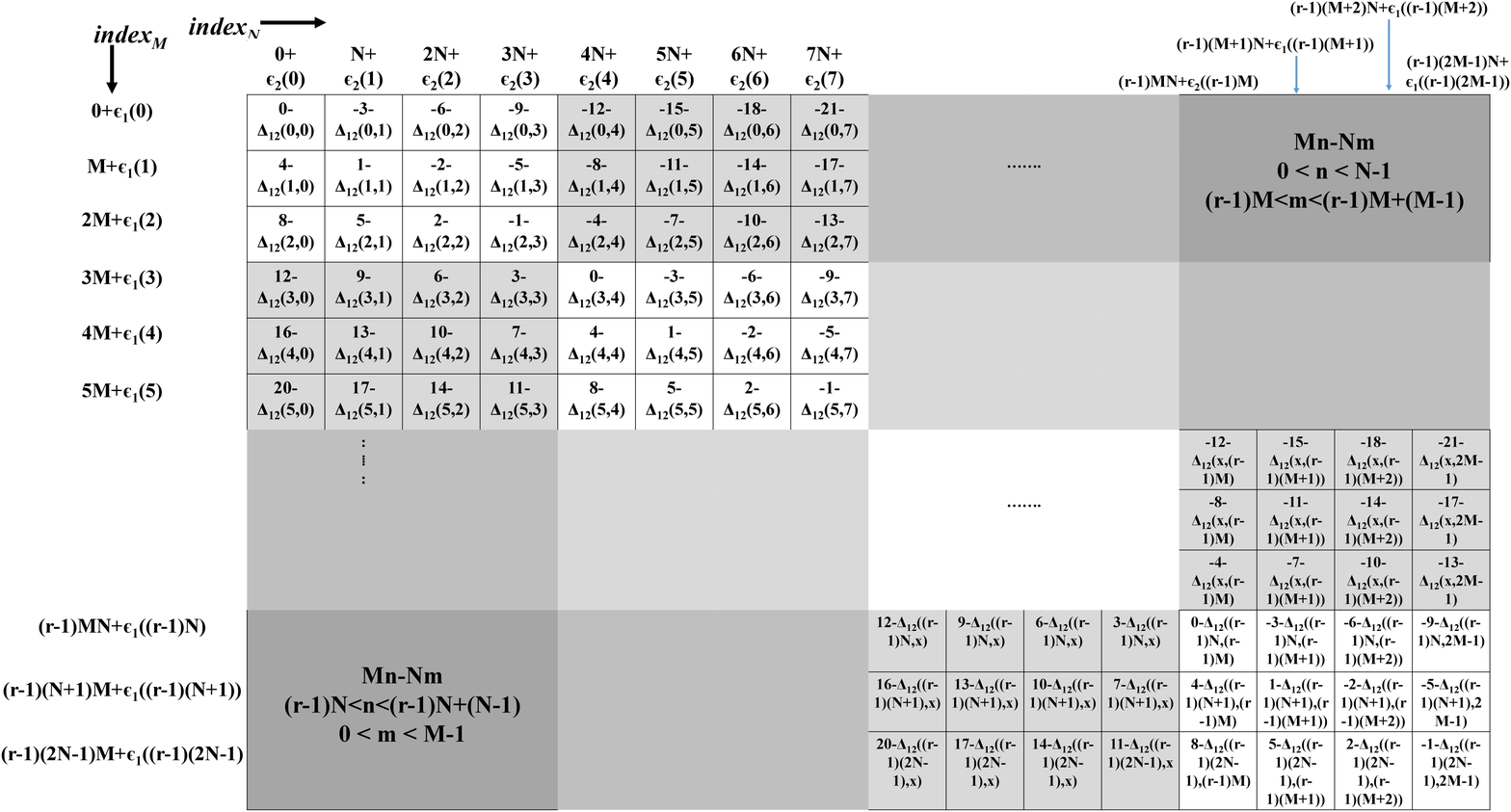}
	\caption{Co-prime sampling with multiple periods: Cross differences in the presence of jitter.}
	\label{jitter_Multi_Lc}
\end{figure*}

Let us now investigate the difference set of a jitter perturbed co-prime sampler with multiple periods. The definitions for self and cross difference sets of the prototype co-prime scheme under the influence of jitter hold true here as well, except that $m\in[0, rM-1]$ and $n\in[0, rN-1]$. Note that $r$ is the number of co-prime periods. These sets are denoted by the same symbols, but with a subscript $r$. 
The union of self differences (positive and negative) of the two sub-samplers is given by:
\begin{eqnarray}\label{eq:self_diffM_chp7}
\nonumber   {\L}^+_{SM_r} \cup {\L}^-_{SM_r} &=& (Mn_1+\epsilon_1(n_1))-(Mn_2+\epsilon_1(n_2))\\
&=& M(n_1-n_2)+\Delta_1(n_1,n_2)
\end{eqnarray}
and,
\begin{eqnarray}\label{eq:self_diffN_chp7}
{\L}^+_{SN_r} \cup {\L}^-_{SN_r} &=& N(m_1-m_2)+\Delta_2(m_1,m_2)
\end{eqnarray}
where $\Delta_1(n_1,n_2)=\epsilon_1(n_1)-\epsilon_1(n_2)$ and $\Delta_2(m_1,m_2)=\epsilon_2(m_1)-\epsilon_2(m_2)$. When $n_1=n_2$ and $m_1=m_2$, $\Delta_1(n_1,n_2)=0$ and $\Delta_2(m_1,m_2)=0$, respectively.
The cross difference set is:
\begin{eqnarray}
\nonumber  {\L}^+_{C_r} &=& (Mn+\epsilon_1(n))-(Nm+\epsilon_2(m))\\
&=& Mn-Nm-\Delta_{12}(n,m)
\end{eqnarray}
and, 
\begin{eqnarray}
{\L}^-_{C_r} &=& Nm-Mn+\Delta_{12}(n,m)
\end{eqnarray}
where $\Delta_{12}(n,m)=\epsilon_2(m)-\epsilon_1(n)$.
The self difference matrix for the individual samplers with multiple periods is shown in Fig.~\ref{jitter_Multi_self_differences}. The cross difference matrix for the set ${\L}^+_{C_r}$ is shown in Fig~\ref{jitter_Multi_Lc}. ${\L}^-_{C_r}$ is a set that contains values which are negative of the values in ${\L}^+_{C_r}$. The number of unique differences in each set of the co-prime sampler with multiple periods under the influence of jitter is given by Proposition~\ref{prop_Multi_jitter_dofs}.
\begin{thm}\label{prop_Multi_jitter_dofs}
	\begin{enumerate}
		\item\label{prop_Multi_dofs_LSM} The sets ${\L}^+_{SM_r}$ and ${\L}^-_{SM_r}$ have a maximum of $\frac{rN(rN-1)}{2}+1$ distinct values.
		\item\label{prop_Multi_dofs_LSN} The sets ${\L}^+_{SN_r}$ and ${\L}^-_{SN_r}$ have a maximum of $\frac{rM(rM-1)}{2}+1$ distinct values.
		\item\label{prop_Multi_dofs_LS+-} The sets ${\L}^+_{S_r}$ and ${\L}^-_{S_r}$ have a maximum of $\frac{rM(rM-1)}{2}+\frac{rN(rN-1)}{2}+1$ distinct values.
		\item\label{prop_Multi_dofs_LS} The set ${\L}_{S_r}$ has a maximum of $rM(rM-1)+rN(rN-1)+1$ distinct values.
		\item\label{prop_Multi_dofs_Lc+-} The sets ${\L}^+_{C_r}$ and ${\L}^-_{C_r}$ have a maximum of $r^2 MN$ distinct values.
		\item\label{prop_Multi_dofs_Lc} The set ${\L}_{C_r}$ has $2r^2 MN$ distinct values.
		\item\label{prop_Multi_dofs_LS_NOTsub_Lc} The self differences ${\L}_{S_r}$ may not be a subset of the cross differences ${\L}_{C_r}$, i.e. ${\L}_{S_r} \nsubseteq {\L}_{C_r}$.
		\item\label{prop_Multi_dofs_L} The set ${\L}={\L}_{C_r} \cup {\L}_{S_r}$ has a maximum of $(rM+rN)(rM+rN-1)+1$ distinct values.
	\end{enumerate}
\end{thm} 
\textit{Proof.}
\begin{enumerate}
	\item As shown in Fig.~\ref{jitter_Multi_LSM}, the number of unique differences in the sets ${\L}^+_{SM_r}$ and ${\L}^-_{SM_r}$ are the unique values in the lower and upper triangle of the self difference matrix respectively, with a common diagonal. This common diagonal has a unique value, i.e. `0', while all the other values in the matrix are different from each other. Therefore, the total number of unique values in the upper triangle as well as the lower triangle is given by:
	\begin{equation}
	\nonumber 1+\sum\limits_{n=1}^{rN-1}n=1+\frac{rN(rN-1)}{2}
	\end{equation}
	For the above equation to hold true $\Delta_1(n_1,n_2)$ should be unique $\forall n_1-n_2=l$, where $1\leq \mid l\mid \leq rN-1$.
	
	\item Similarly, from Fig.~\ref{jitter_Multi_LSN} the number of unique values in the sets ${\L}^+_{SN_r}$ and ${\L}^-_{SN_r}$ can be written as:
	\begin{equation}
	\nonumber 1+\sum\limits_{m=1}^{rM-1}m=1+\frac{rM(rM-1)}{2}
	\end{equation}
	For the above equation to hold true $\Delta_2(m_1,m_2)$ should be unique $\forall m_1-m_2=l$, where $1\leq \mid l\mid \leq rM-1$.
	
	\item ${\L}^+_{SM_r}$ and ${\L}^+_{SN_r}$ (${\L}^-_{SM_r}$ and ${\L}^-_{SN_r}$) have `0' as a common value, hence  ${\L}^+_{S_r}$ (${\L}^-_{S_r}$) has $\frac{rM(rM-1)}{2}+\frac{rN(rN-1)}{2}+1$ unique values.
	
	\item The only overlapping self difference value between sampler $x(Mn+\epsilon_1(n))$ and $x(Nm+\epsilon_2(m))$ i.e. (${\L}^+_{SM_r} \cup  {\L}^-_{SM_r}$) and (${\L}^+_{SN_r} \cup  {\L}^-_{SN_r}$) is `0'. Hence, the unique values in set ${\L}_{S_r}={\L}^+_{SM_r} \cup  {\L}^+_{SN_r} \cup {\L}^-_{SM_r} \cup  {\L}^-_{SN_r}$ is:
	\begin{equation}
	\begin{split}
	\nonumber   2\left(\frac{rM(rM-1)}{2}+\frac{rN(rN-1)}{2}+1\right)-1\\
	=rM(rM-1)+rN(rN-1)+1
	\end{split}
	\end{equation}
	Thus justifying the Proposition~\ref{prop_Multi_jitter_dofs}-\ref{prop_Multi_dofs_LS}.
	\item Since the total number of elements in the set ${\L}^+_{C_r}$ is $r^2MN$, we need to show that these elements are unique. Let $l_{c_1}=Mn_1+\epsilon_1(n_1)-(Nm_1+\epsilon_2(m_1))$ and $l_{c_2}=Mn_2+\epsilon_1(n_2)-(Nm_2+\epsilon_2(m_2))$ be the elements in the jitter perturbed set ${\L}^+_{C_r}$. Let us assume that $l_{c_1}=l_{c_2}$ for some $0\leq n_1,n_2\leq rN-1$ and $0\leq m_1,m_2\leq rM-1$, then we have:
	\begin{align}
	\nonumber   Mn_1+\epsilon_1(n_1)-(Nm_1+\epsilon_2(m_1))=Mn_2+\epsilon_1(n_2)\\
	\nonumber   -(Nm_2+\epsilon_2(m_2))
	\end{align}
	\begin{align}
	\nonumber    M(n_1-n_2)&=&N(m_1-m_2)-(\epsilon_1(n_1)-\epsilon_1(n_2))\\
	\nonumber    &&+(\epsilon_2(m_1)-\epsilon_2(m_2))\\
	\frac{M}{N}&=&\frac{(m_1-m_2)+\frac{\Delta_2(m_1,m_2)-\Delta_1(n_1,n_2)}{N}}{n_1-n_2}
	\end{align}
	Since $\epsilon_1(n_1)$, $\epsilon_1(n_2)$, $\epsilon_2(m_1)$, and $\epsilon_2(m_2)$ take values in the range $(-\frac{1}{4},\frac{1}{4})$, we have $\Delta_1(n_1,n_2)$ and $\Delta_2(m_1,m_2)$ in the range $(-\frac{1}{2},\frac{1}{2})$. Using the extreme values of this range, it is easy to obtain the range for $\Delta_2(m_1,m_2)-\Delta_1(n_1,n_2)$ as $(-1,1)$. 
	When $\Delta_2(m_1,m_2)-\Delta_1(n_1,n_2)=0$, we have:
	\begin{equation}\label{Proof_Prop7.4_5_1}
	\frac{M}{N}=\frac{(m_1-m_2)+0}{n_1-n_2}
	\end{equation}
	Since $(M,N)$ are co-prime, $m_1-m_2<rM$ and $n_1-n_2<rN$, which implies that equation~\eqref{Proof_Prop7.4_5_1} can hold and hence the proposition fails. But when $\Delta_2(m_1,m_2)-\Delta_1(n_1,n_2)=\pm 1$, we have:
	\begin{equation}\label{Proof_Prop7.4_5_2}
	\frac{M}{N}=\frac{(m_1-m_2)\pm \frac{1}{N}}{n_1-n_2}
	\end{equation}
	Since $\frac{1}{N}<<1$ and is not an integer, it cannot produce the co-prime ratio on the right-hand-side of~\eqref{Proof_Prop7.4_5_2}, hence, the proposition holds. In fact it holds for any value of $\Delta_2(m_1,m_2)-\Delta_1(n_1,n_2)$ in the range $\pm(0,1]$, which excludes zero. Hence, set ${\L}^+_{C_r}$ has $r^2MN$ unique differences. A similar argument holds for set ${\L}^-_{C_r}$, thus proving Proposition~\ref{prop_Multi_jitter_dofs}-\ref{prop_Multi_dofs_Lc+-}. The condition for this proposition to hold is given below:\\
	\begin{eqnarray}\label{Proof_Prop7.4_5_multi}
	\nonumber \Delta_2(m_1,m_2)-\Delta_1(n_1,n_2)\not= 0\\
	\nonumber (\epsilon_2(m_1)-\epsilon_2(m_2))-(\epsilon_1(n_1)-\epsilon_1(n_2))\not= 0\\
	\nonumber \epsilon_2(m_1)-\epsilon_1(n_1)\not= \epsilon_2(m_2))-\epsilon_1(n_2))\\
	\Delta_{12}(n_1,m_1)\not= \Delta_{12}(n_2,m_2)
	\end{eqnarray}
	This implies that the proposition holds provided the jitters in the samples are such that their cross differences (only of the jitter values) are unique.
	\item Let $l_{c_1}=Mn_1+\epsilon_1(n_1)-(Nm_1+\epsilon_2(m_1))$ and $l_{c_2}=Nm_2+\epsilon_2(m_2)-(Mn_2+\epsilon_1(n_2))$ be the elements in the jitter perturbed set ${\L}^+_{C_r}$ and ${\L}^-_{C_r}$ respectively. Let us assume that $l_{c_1}=l_{c_2}$ for some $0\leq n_1,n_2\leq rN-1$ and $0\leq m_1,m_2\leq rM-1$, then
	\begin{align}
	\nonumber Mn_1+\epsilon_1(n_1)-(Nm_1+\epsilon_2(m_1))=Nm_2+\epsilon_2(m_2)\\
	\nonumber -(Mn_2+\epsilon_1(n_2))
	\end{align}
	\begin{align}   
	\nonumber    M(n_1+n_2)=N(m_1+m_2)+(\epsilon_2(m_1)-\epsilon_1(n_1))\\
	\nonumber   +(\epsilon_2(m_2)-\epsilon_1(n_2))\\
	\frac{M}{N}=\frac{(m_1+m_2)+\frac{\Delta_{12}(n_1,m_1)+\Delta_{12}(n_2,m_2)}{N}}{n_1+n_2}
	\end{align}
	Since $\epsilon_1(n_1)$, $\epsilon_1(n_2)$, $\epsilon_2(m_1)$, and $\epsilon_2(m_2)$ take values in the range $(-\frac{1}{4},\frac{1}{4})$, we have $\Delta_{12}(n_1,m_1)$ and $\Delta_{12}(n_2,m_2)$ in the range $(-\frac{1}{2},\frac{1}{2})$. Using the extreme values of this range, it is easy to obtain the range for $\Delta_{12}(n_1,m_1)$ $+\Delta_{12}(n_2,m_2)$ as $(-1,1)$. 
	When $\Delta_{12}(n_1,m_1)$ $+\Delta_{12}(n_2,m_2)$ $=0$, we have:
	\begin{equation}\label{Proof_Prop7.4_6_1}
	\frac{M}{N}=\frac{(m_1+m_2)+0}{n_1+n_2}
	\end{equation}
	Since $m_1+m_2<2rM$ and $n_1+n_2<2rN$, there is a possibility of obtaining the co-prime ratio on the right-hand-side of equation~\eqref{Proof_Prop7.4_6_1}. Hence, the Proposition~\ref{prop_Multi_dofs_Lc} fails. But under the assumption that $\Delta_{12}(n_1,m_1)$ $+\Delta_{12}(n_2,m_2)$ $\neq 0$, it holds. Let us assume that $\Delta_{12}(n_1,m_1)$ $+\Delta_{12}(n_2,m_2)$ takes an extreme value of $\pm 1$. 
	\begin{equation}\label{Proof_Prop7.4_6_2}
	\frac{M}{N}=\frac{(m_1+m_2)\pm \frac{1}{N}}{n_1+n_2}
	\end{equation}
	Since $m_1+m_2+\frac{1}{N}$ is not an integer, we cannot produce the co-prime ratio on the right-hand-side of~\eqref{Proof_Prop7.4_6_2}. In fact it holds for any value of $\Delta_{12}(n_1,m_1)+\Delta_{12}(n_2,m_2)$ in the range $\pm(0,1]$, which excludes zero. This implies that:
	\begin{eqnarray}\label{Proof_Prop7.4_6_multi}
	\nonumber  \Delta_{12}(n_1,m_1)+\Delta_{12}(n_2,m_2)\not = 0\\
	\Delta_{12}(n_1,m_1)\not = -\Delta_{12}(n_2,m_2)
	\end{eqnarray}
	Since sets ${\L}^+_{C_r}$ and ${\L}^-_{C_r}$ have $r^2MN$ unique differences, it can be safely concluded that ${\L}={\L}^+_{C_r} \cup {\L}^-_{C_r}$ has $2r^2MN$ unique values provided both equations~\eqref{Proof_Prop7.4_5_multi} and~\eqref{Proof_Prop7.4_6_multi} are satisfied. This implies:
	\begin{eqnarray}\label{Proof_Prop7.4_6_multi_condition}
	\nonumber |\Delta_{12}(n_1,m_1)| \neq |\Delta_{12}(n_2,m_2)|
	\end{eqnarray}
	It may be noted that the conditions derived for the validity of Proposition~\ref{prop_Multi_jitter_dofs}-\ref{prop_Multi_dofs_Lc+-} and~\ref{prop_Multi_jitter_dofs}-\ref{prop_Multi_dofs_Lc}, are sufficient conditions but not necessary conditions. 
	For Proposition~\ref{prop_Multi_jitter_dofs}-\ref{prop_Multi_dofs_Lc+-}, the necessary condition is:
	\begin{align}\label{Proof_Prop7.4_5_multi_necessary_condn}
	\nonumber \Delta_{12}(n_1,m_1) \neq \Delta_{12}(n_2,m_2),\\
	\forall \left\{m_1,m_2,n_1,n_2 \big| \frac{m_1-m_2}{n_1-n_2}=\frac{M}{N}\right\}
	\end{align}
	For Proposition~\ref{prop_Multi_jitter_dofs}-\ref{prop_Multi_dofs_Lc}, the necessary condition is:
	\begin{align}\label{Proof_Prop7.4_6_multi_necessary_condn}
	\nonumber |\Delta_{12}(n_1,m_1)| \neq |\Delta_{12}(n_2,m_2)|,\\
	\forall \left\{m_1,m_2,n_1,n_2 \bigm| \frac{m_1+m_2}{n_1+n_2}=\frac{M}{N}\right\}
	\end{align}
	\item The proof for Proposition~\ref{prop_Multi_jitter_dofs}-\ref{prop_Multi_dofs_LS_NOTsub_Lc} is same as for Proposition I-(7) in~\cite{U_S_2} and is given below:\\
	Let $l_{c}=Mn+\epsilon_1(n)-(Nm+\epsilon_2(m))$ be an element in the perturbed set ${\L}^+_{C_r}$. Substituting $m=0$ in this equation leads to;
	\begin{equation}\label{eq:Proof_Prop7.4_7_1}
	l_c=Mn-\Delta_{12}(n,0)
	\end{equation}
	Letting $n_2=0$ and $n_1=n$ in the self difference equation~\eqref{eq:self_diffM_chp7} gives
	\begin{equation}\label{eq:Proof_Prop7.4_7_2}
	l_s=Mn+\Delta_1(n,0)
	\end{equation}
	Equations \eqref{eq:Proof_Prop7.4_7_1} and \eqref{eq:Proof_Prop7.4_7_2} are not equal under the assumption that $-\Delta_{12}(n,0)\neq \Delta_1(n,0)$.
	Next, substitute $n=0$ in the equation for cross difference $l_c$ which leads to:
	\begin{equation}\label{eq:Proof_Prop7.4_7_3}
	l_c=-Nm-\Delta_{12}(0,m)
	\end{equation}
	Letting $m_1=0$ and $m_2=m$ in the self difference equation~\eqref{eq:self_diffN_chp7} gives:
	\begin{equation}\label{eq:Proof_Prop7.4_7_4}
	l_s=-Nm+\Delta_2(0,m)
	\end{equation}
	Equations \eqref{eq:Proof_Prop7.4_7_3} and \eqref{eq:Proof_Prop7.4_7_4} are not equal under the assumption that $-\Delta_{12}(0,m)\neq \Delta_2(0,m)$. Similarly, one can argue for $l_c \in {\L}^-_{C_r}$. Therefore, proving Proposition~\ref{prop_Multi_jitter_dofs}-\ref{prop_Multi_dofs_LS_NOTsub_Lc}.
	\item The combined set ${\L}={\L}_{C_r} \cup {\L}_{S_r}={\L}^+_{C_r} \cup {\L}^-_{C_r} \cup {\L}^+_{S_r} \cup {\L}^-_{S_r}$. From Proposition~\ref{prop_Multi_jitter_dofs}-\ref{prop_Multi_dofs_LS_NOTsub_Lc}, the number of distinct values in ${\L}$ is the sum of the unique values in ${\L}_{S_r}$ (Proposition~\ref{prop_Multi_jitter_dofs}-\ref{prop_Multi_dofs_LS}) and ${\L}_{C_r}$ (Proposition~\ref{prop_Multi_jitter_dofs}-\ref{prop_Multi_dofs_Lc}), and is given below:
	\begin{eqnarray}
	\nonumber &&2r^2MN+rM(rM-1)+rN(rN-1)+1\\
	\nonumber &=&rM(rN+rM-1)+rN(rM+rN-1)+1\\
	\nonumber &=&(rM+rN)(rM+rN-1)+1
	\end{eqnarray}    
\end{enumerate}
\subsection{Weight function under the influence of Jitter}
\label{JITTER_multi_weight_fn}
The weight function for the co-prime samplers with multiple periods under the influence of sampling jitters is given by Proposition~\ref{prop_Multi_jitter_wts}, and is similar to Proposition II in~\cite{U_S_2} which was derived for the prototype co-prime samplers with perturbations.
The work in~\cite{U_S_2} had described two systems; a blind system and a non-blind system, and is also considered here for the multiple period scenario. The blind system is a system in which the presence of jitter is unknown and hence it follows the procedure used in the ideal scenario for autocorrelation estimation. The weight function in this case, after mapping $[l-\frac{1}{2}, l+\frac{1}{2})\rightarrow l$, was shown to be the same as that of the prototype co-prime array without jitters (refer Fig. 4 and Fig. 6(b) in~\cite{U_S_2}). The co-prime samplers with multiple periods have sample indices $cMN$ (where $0\leq c\leq r-1$) coinciding under ideal conditions. This does not hold true in the presence of jitter. However, the blind system for the multiple period scenario assumes that the samples acquired at these indices are the same and hence uses only one of them (either $x(Mn)|n=cN$ or $x(Nm)|m=cM$) in the combined set for estimation. Therefore, the weight function in this case is the same as that of the co-prime arrays with multiple periods~\cite{UVD_phdthesis}. On the other hand, we have a non-blind system which efficiently utilizes the information available in the data for estimation in the presence of jitters.

Let $z_r(l)$ represent the number of elements available for autocorrelation estimation at value $l$ for co-prime samplers with multiple periods. $l$ represents the unmapped location and may not be an integer.

\begin{thm}\label{prop_Multi_jitter_wts}
	\begin{enumerate}
		\item\label{prop_Multi_jitter_wts_Lc+} For $l \in {\L}^+_{C_r}$, $z_r(l)=1$
		\item\label{prop_Multi_jitter_wts_Lc-} For $l \in {\L}^-_{C_r}$, $z_r(l)=1$
		\item\label{prop_Multi_jitter_wts_Lc} For $l \in {\L}_{C_r}={\L}^+_{C_r} \cup {\L}^-_{C_r}$, $z_r(l)=1$
		\item\label{prop_Multi_jitter_wts_L0} For $l=0$, $z_r(l)=rM+rN$
		\item\label{prop_Multi_jitter_wts_LSM} For $l \in {\L}^+_{SM_r} \cup {\L}^-_{SM_r}-\{0\}$, $z_r(l)=1$
		\item\label{prop_Multi_jitter_wts_LSN} For $l \in {\L}^+_{SN_r} \cup {\L}^-_{SN_r}-\{0\}$, $z_r(l)=1$
	\end{enumerate}
\end{thm} 
These claims are based on Proposition~\ref{prop_Multi_jitter_dofs}, which described the number of unique differences in each set. Hence the assumptions made in Proposition~\ref{prop_Multi_jitter_dofs} also hold true for Proposition~\ref{prop_Multi_jitter_wts}.

The non-blind system seeks to improve the number of unique sample pairs for autocorrelation estimation in the presence of sampling jitter and is given by Proposition~\ref{prop_Multi_jitter_wts_mapped} after mapping the differences in the range $l\pm\frac{1}{2}\rightarrow l$. Let $z_{nb_r}(l)$ represent the number of available contributors for autocorrelation estimation at $l$, for the non-blind system. Here, $l$ is an integer and represents the mapped locations.
\begin{thm}\label{prop_Multi_jitter_wts_mapped}
	\begin{enumerate}
		\item\label{prop_Multi_jitter_wts_mapped_LcMN} For difference values that are multiples of $MN$:
		\begin{align}
		\nonumber        z_{nb_r}(l)=rM+rN+r,~\text{for}~\{l=0\}
		\end{align}
		\begin{align}
		\nonumber        z_{nb_r}(l)=(r-c)M+(r-c)N+2(r-c)\\
		\nonumber        \text{for}~\{l=\pm cMN, 1<c<(r-1)\}
		\end{align}
		\item\label{prop_Multi_jitter_wts_mapped_LSM} For $l$ belonging to the set ${\L}^+_{SM_r} \cup {\L}^-_{SM_r}$ excluding the difference values that are multiples of $MN$:
		\begin{align}
		\nonumber        z_{nb_r}(l)=(rN-i)+(r-\left\lfloor \frac{i}{N}\right\rfloor)+(r-\left\lceil \frac{i}{N}\right\rceil)\\
		\nonumber        \text{for}~\{1\leq i \leq rN-1,\frac{i}{N}\not\in \mathbb{Z}, l=\pm Mi\}
		\end{align}
		\item\label{prop_Multi_jitter_wts_mapped_LSN} For $l$ belonging to the set ${\L}^+_{SN_r} \cup {\L}^-_{SN_r}$ excluding the difference values that are multiples of $MN$:
		\begin{align}
		\nonumber        z_{nb_r}(l)=(rM-i)+(r-\left\lfloor\frac{i}{M}\right\rfloor)+(r-\left\lceil\frac{i}{M}\right\rceil)\\
		\nonumber        \text{for}~\{1\leq i \leq rM-1,\frac{i}{M}\not\in \mathbb{Z}, l=\pm Ni\}
		\end{align}
		\item\label{prop_Multi_jitter_wts_mapped_LC-S} For $l$ belonging to the set ${\L}_{C_r} -{\L}_{S_r}$, i.e. cross differences without any self difference value:
		\begin{align}
		\nonumber  z_{nb_r}(l)=2r,~\text{for}~\{l \in {\L}_{C_r}-{\L}_{S_r},\\
		\nonumber  0\leq\mid l\mid\leq MN-M-N\}
		\end{align}
		\begin{align}
		\nonumber  z_{nb_r}(l)=2(r-i),\\
		~\text{for}~\{l \in {\L}_{C_r}-{\L}_{S_r},(iN+1)M-(M-1)N\\
		\nonumber  \leq \mid l\mid \leq ((i+1)N-1)M-N,i\in[1,r-1]\}
		\end{align}
	\end{enumerate}
\end{thm}
\begin{figure*}
	\centering
	\includegraphics[width=0.75\textwidth]{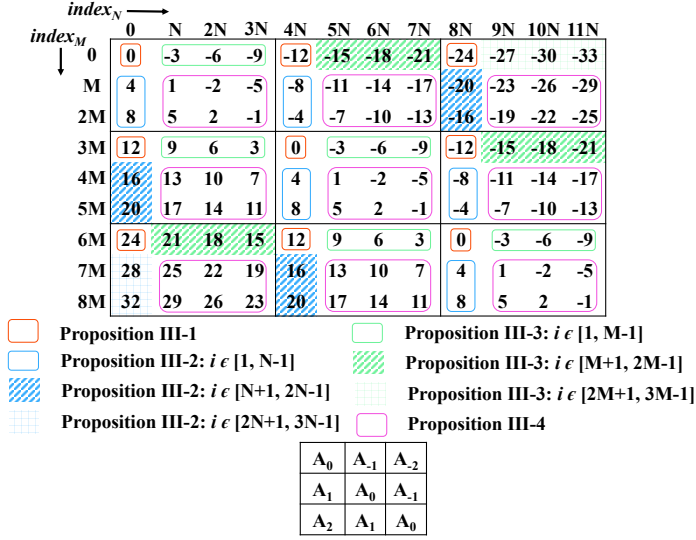}
	\caption{Cross difference set: $M=4$, $N=3$ and $r=3$.}
	\label{fig:jitter_Multi_Lc_proposition_non_blind}
\end{figure*}
Since the self differences are not a subset of the cross differences, the weight function of the non-blind system has additional unique sample pairs mapped to the self differences from the set ${\L}^+_{C_r}$. For a better understanding of Proposition~\ref{prop_Multi_jitter_wts_mapped}, as an example, we consider a cross difference set with $M=4$, $N=3$ and $r=3$ as shown in Fig.~\ref{fig:jitter_Multi_Lc_proposition_non_blind}.
For the case when $l=0$, the self differences have $rM+rN$ unique sample pairs as in Fig.~\ref{jitter_Multi_self_differences} ($\Delta_1(cN, cN)\not=\Delta_2(cM, cM)$) plus additional pairs from the set ${\L}^+_{C_r}$ under the assumption that $\Delta_{12}(c_{1}N, c_{1}M)\not=\Delta_{12}(c_{2}N, c_{2}M)$ $\forall~0\leq c_{1}, c_{2}\leq r-1$. Thus leading to $rM+rN+r$ contributors. For the case when $l=MN$, i.e. 4x3=12, the cross difference set has 12 appearing $(r-1)$ times in the region $A_{1}$ and $-12$ appears $(r-1)$ times in the region $A_{-1}$. They are generated by a unique pair of indices and are not contained in the self difference set under the assumptions made in Proposition~\ref{prop_Multi_jitter_dofs}. Therefore, under the wide sense stationary condition, the number of contributors at $l=12$ is $(r-1)M+(r-1)N+2(r-1)$. In general, for $l=cMN$ where $c\in[1, r-1]$ the expression is given by $(r-c)M+(r-c)N+2(r-c)$ and justifies Proposition~\ref{prop_Multi_jitter_wts_mapped}-\ref{prop_Multi_jitter_wts_mapped_LcMN}.

For the case when $l=\pm Mi$, where $i\in [0, rN-1]$, and excluding $l=cMN$, the number of contributors per difference value is given by $(rN-n)$ from the self difference matrix (Fig.~\ref{jitter_Multi_LSM}) plus the contributors from the cross difference set as shown in Fig.~\ref{fig:jitter_Multi_Lc_proposition_non_blind}. If $i\in [1, N-1]$ i.e. $l=\{4, 8\}$, we have $r+(r-1)=5$ additional contributors. If $i\in [N+1, 2N-1]$ i.e. $l=\{16, 20\}$ and $i\in [2N+1, 3N-1]$ i.e. $l=\{28, 32\}$ we have $(r-1)+(r-2)$ and $(r-2)+(r-3)$ additional contributors respectively. Thus justifying Proposition~\ref{prop_Multi_jitter_wts_mapped}-\ref{prop_Multi_jitter_wts_mapped_LSM}. A similar argument holds for Proposition~\ref{prop_Multi_jitter_wts_mapped}-\ref{prop_Multi_jitter_wts_mapped_LSN}.
The difference values in the set ${\L}_{C_r}-{\L}_{S_r}$ appear $\{2r, 2(r-1), ..., 2\}$ times in $\{A_0, A_1+A_{-1}, ..., A_{r-1}+A_{-(r-1)}\}$ regions respectively under the assumption that the signal is wide sense stationary (refer Fig.~\ref{fig:jitter_Multi_Lc_proposition_non_blind}). This justifies the claims made in Proposition~\ref{prop_Multi_jitter_wts_mapped}-\ref{prop_Multi_jitter_wts_mapped_LC-S}.

\begin{figure*}[!t]
	\centering
	\subfloat[$r=1$]{\includegraphics[width=0.48\textwidth]{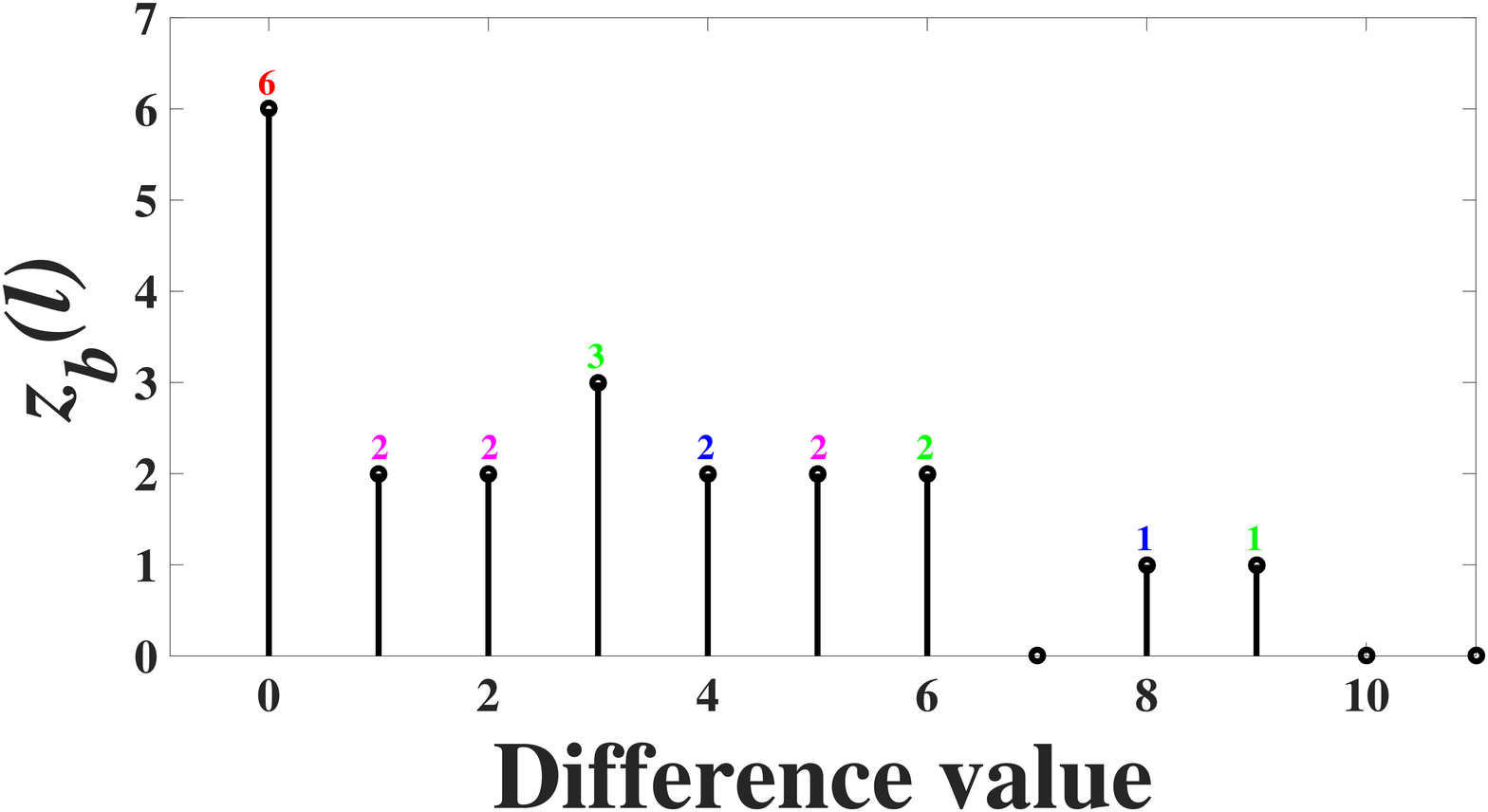}%
		\label{Blind_wts_1}}
	\hfil
	\subfloat[$r=2$]{\includegraphics[width=0.48\textwidth]{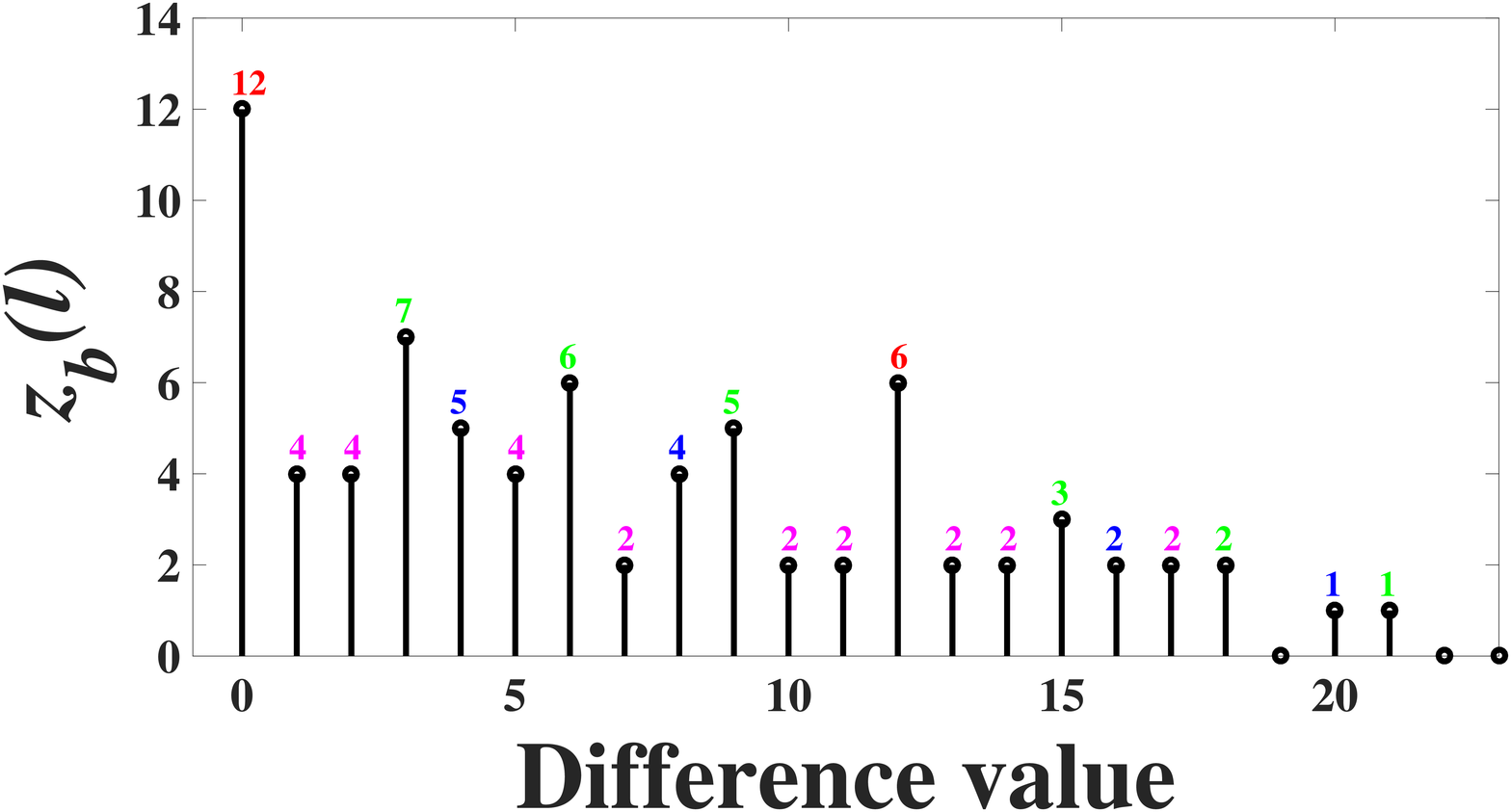}%
		\label{Blind_wts_2}}
	\hfil
	\subfloat[$r=3$]{\includegraphics[width=0.48\textwidth]{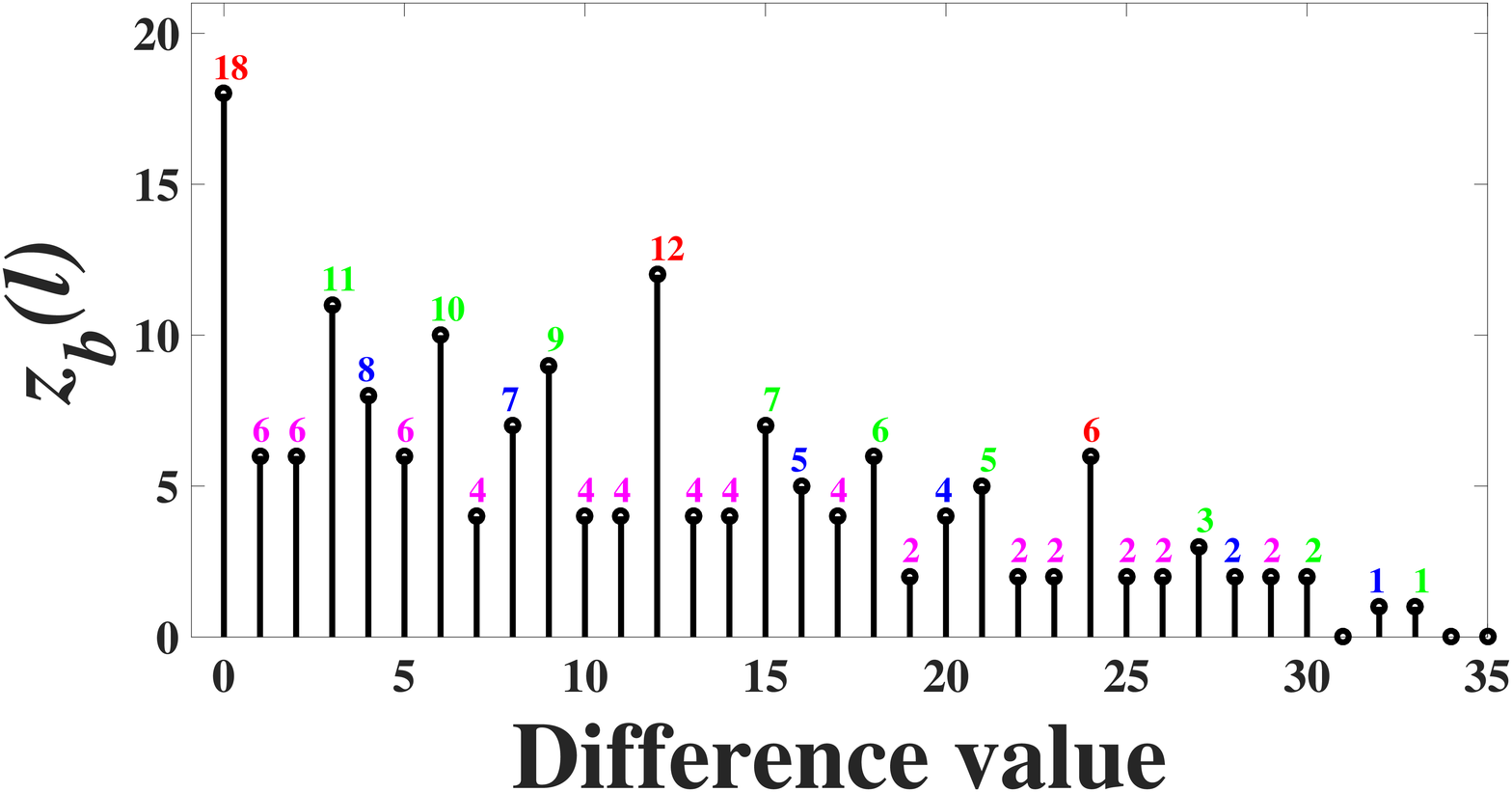}%
		\label{Blind_wts_3}}
	\hfil
	\subfloat[$r=4$]{\includegraphics[width=0.48\textwidth]{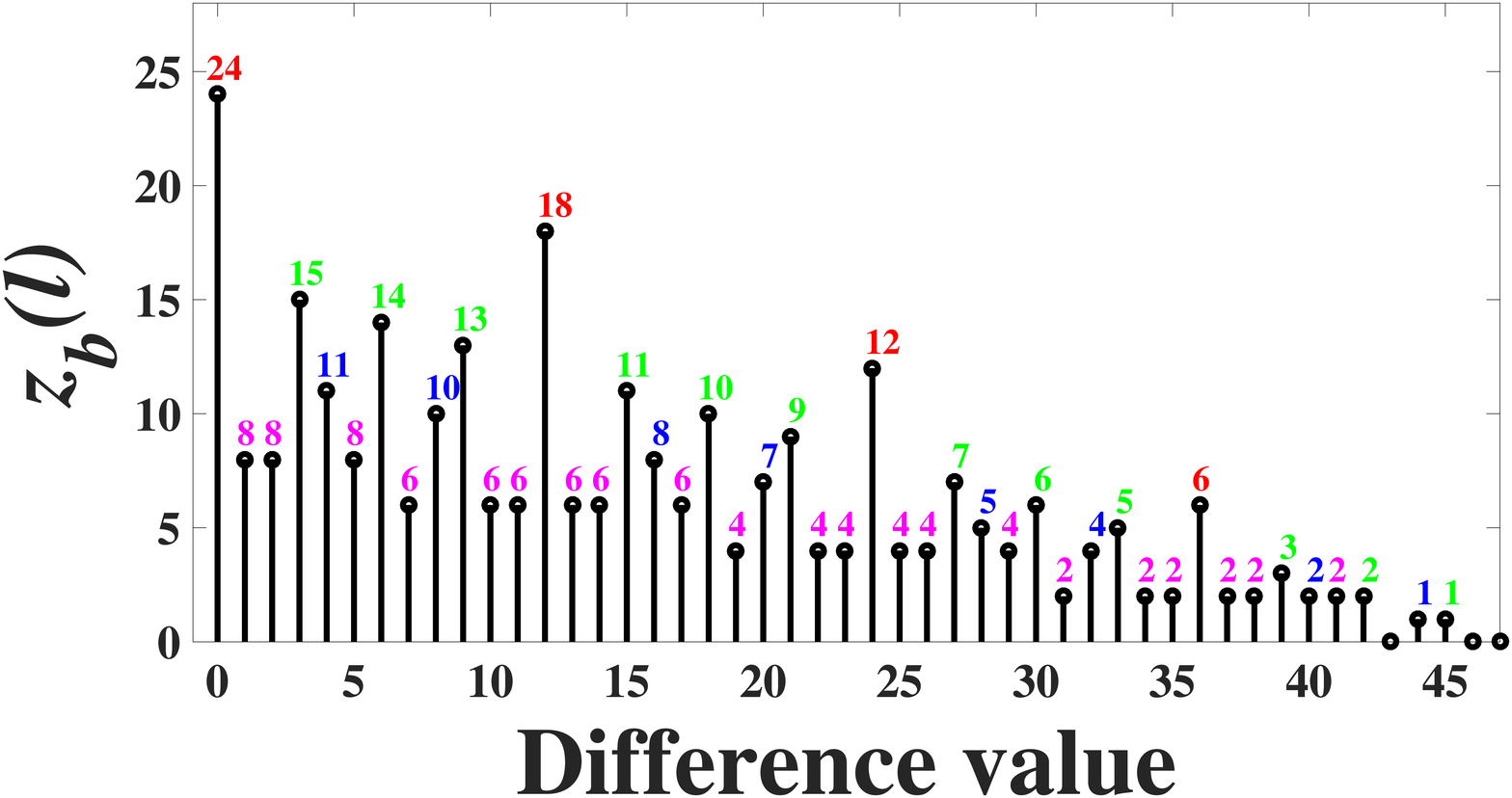}%
		\label{Blind_wts_4}}
	\caption{Number of contributors for a blind system.}
	\label{fig:contibutors_blind}
\end{figure*}

\begin{figure*}[!t]
	\centering
	\subfloat[$r=1$]{\includegraphics[width=0.48\textwidth]{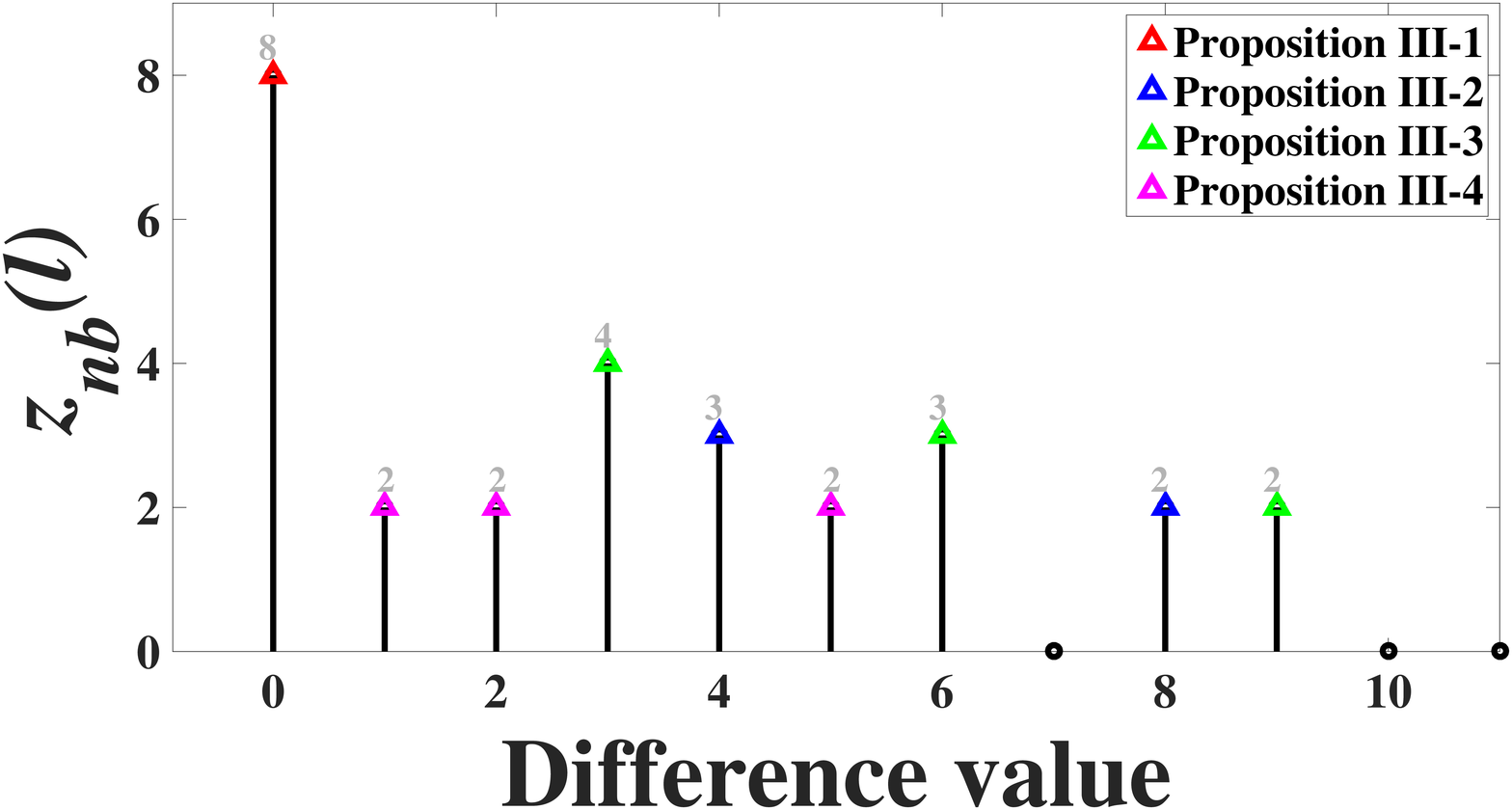}%
		\label{Non_Blind_wts_1}}
	\hfil
	\subfloat[$r=2$]{\includegraphics[width=0.48\textwidth]{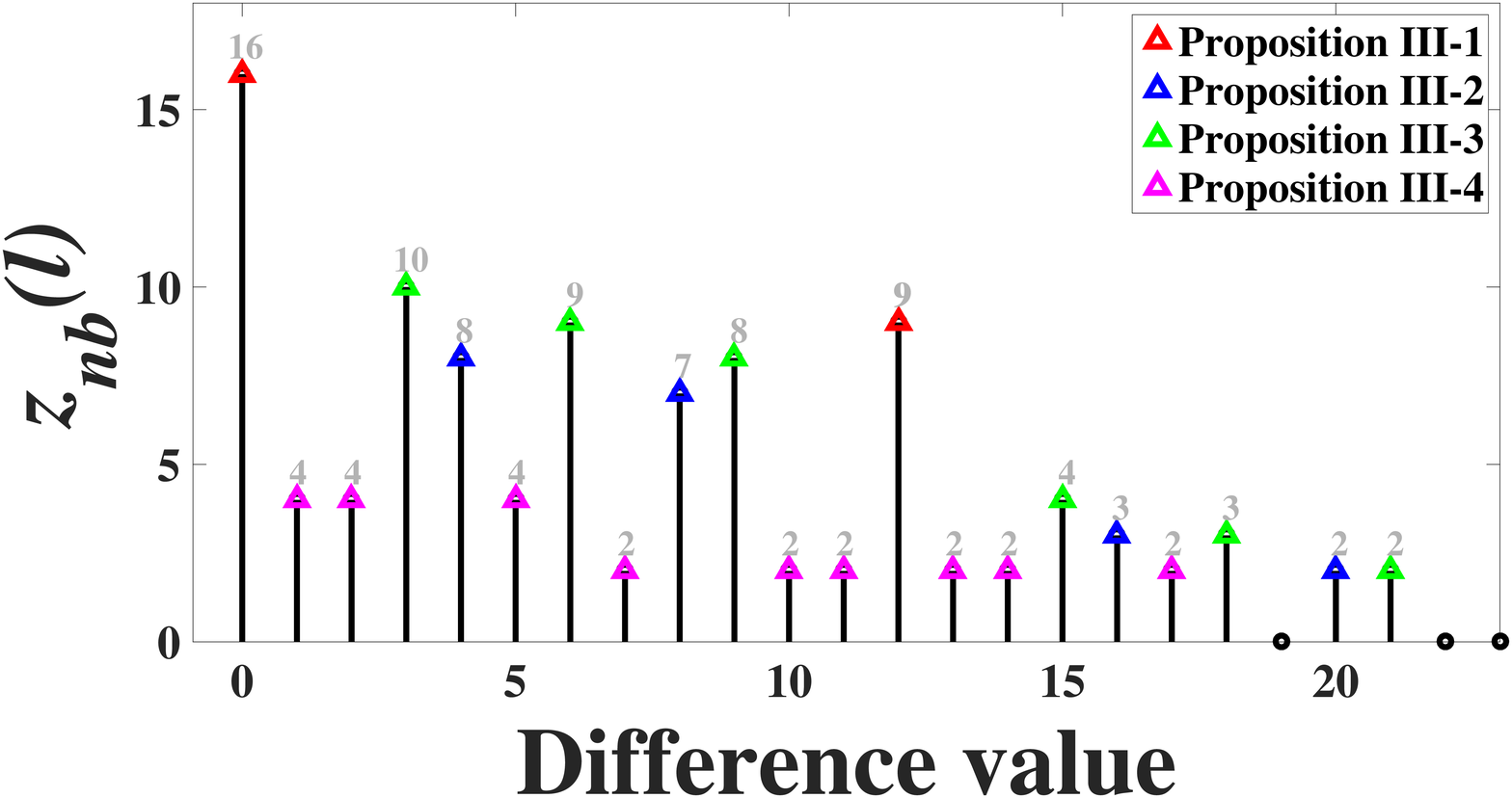}%
		\label{Non_Blind_wts_2}}
	\hfil
	\subfloat[$r=3$]{\includegraphics[width=0.48\textwidth]{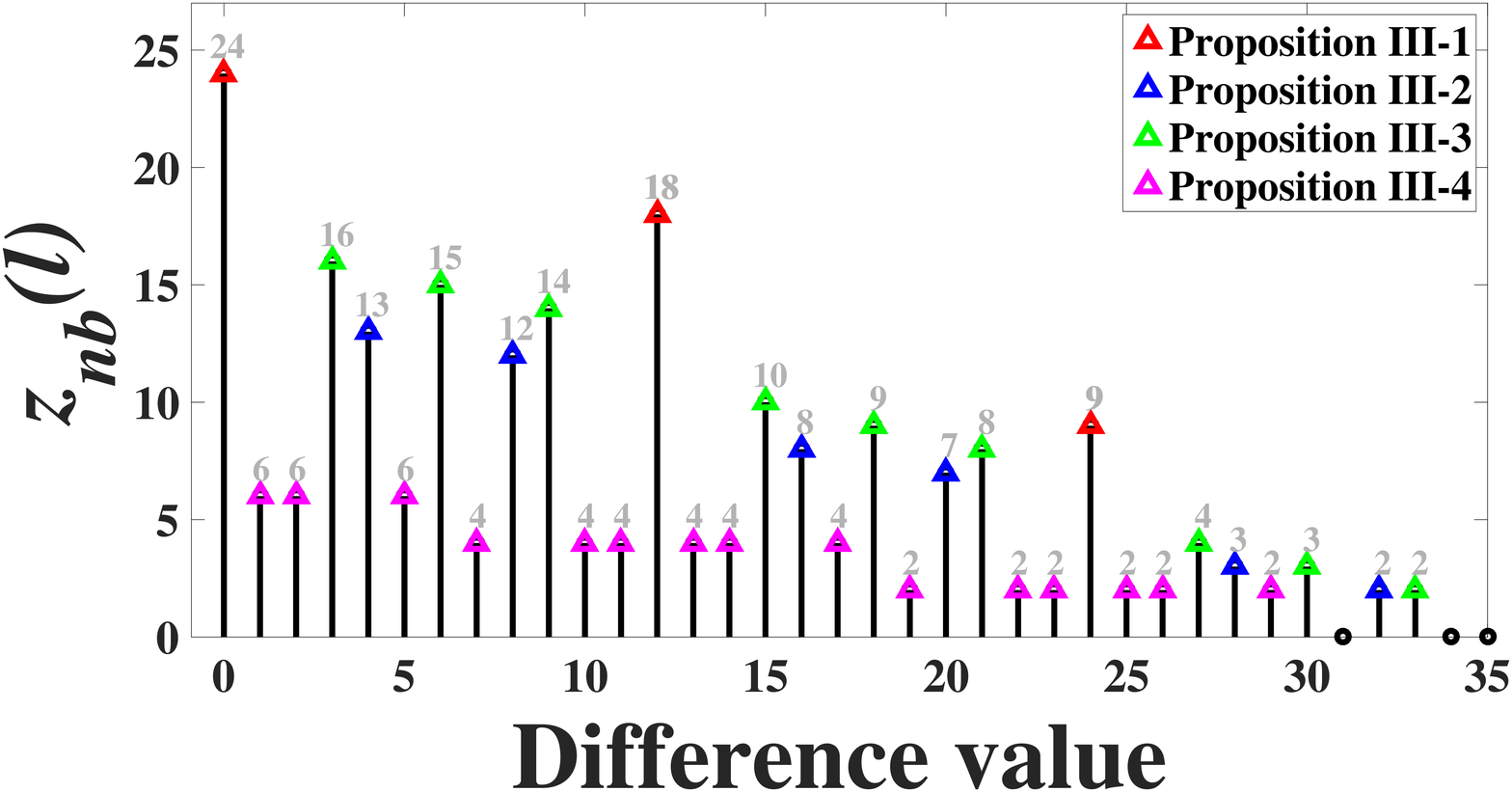}%
		\label{Non_Blind_wts_3}}
	\hfil
	\subfloat[$r=4$]{\includegraphics[width=0.48\textwidth]{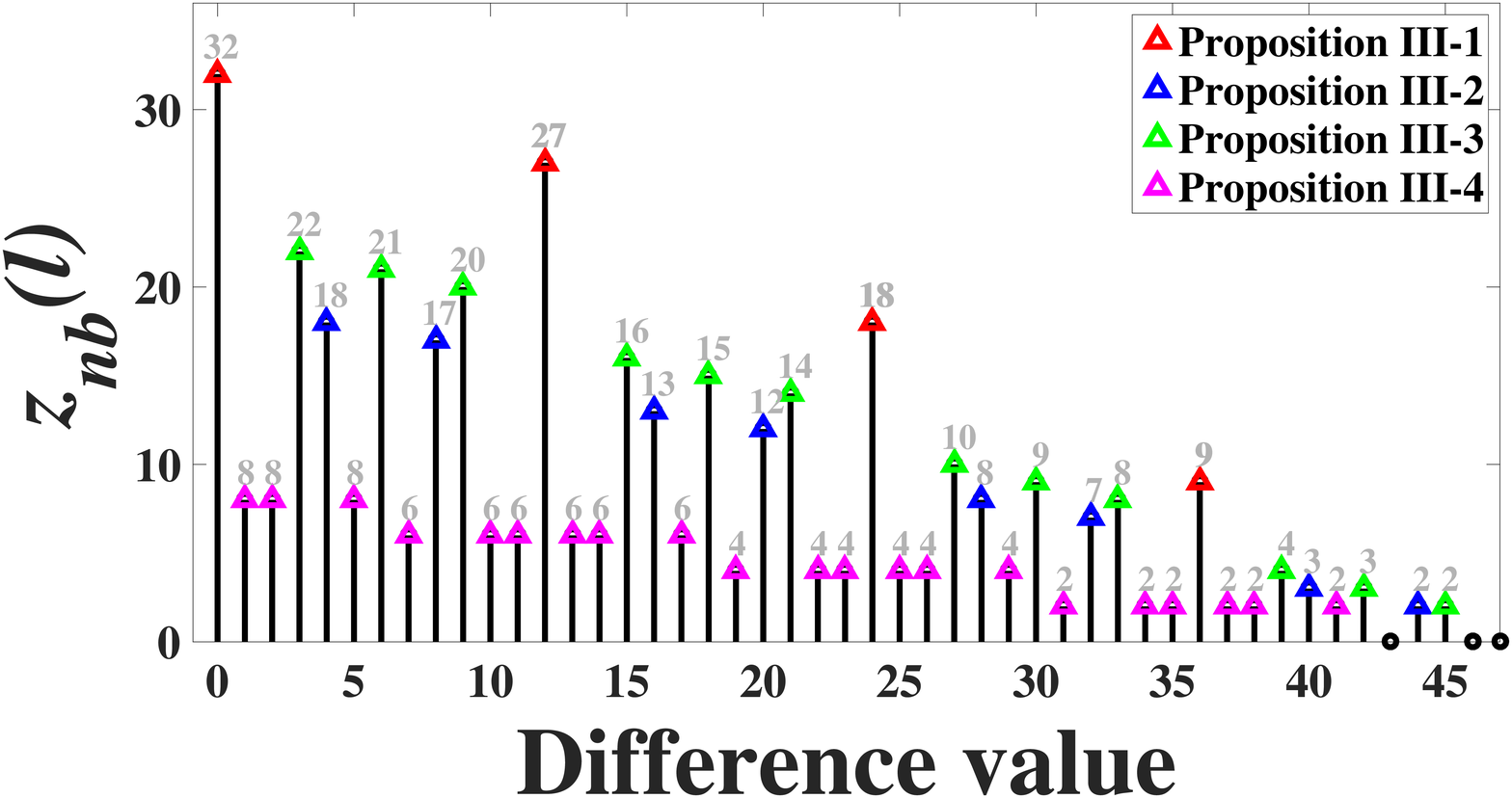}%
		\label{Non_Blind_wts_4}}
	\caption{Number of contributors for a non-blind system.}
	\label{fig:contibutors_non_blind}
\end{figure*}

The number of contributors for autocorrelation estimation for a blind and a non-blind co-prime sampler with multiple periods is shown in Fig.~\ref{fig:contibutors_blind} and~\ref{fig:contibutors_non_blind} with $r=\{1, 2, 3, 4\}$. It may be noted that the weight function for a blind system after mapping is the same as the ideal multiple period weight function. It is evident that the non-blind system has more number of contributors for autocorrelation estimation and increases as a function of $r$. The additional contributors available is given by:
\begin{align}\label{eq:jitter_multi_additional_contributors}
\nonumber    2r+\{3(r-1)+3(r-2)+ ... +3\}\\
\nonumber    +\sum\limits_{r_j=1}^{r}\sum\limits_{i=1}^{N-1}\{(r_j-\lfloor \frac{i}{N}\rfloor)+(r_j-\lceil \frac{i}{N}\rceil)\}\\
\nonumber    +\sum\limits_{r_j=1}^{r}\sum\limits_{i=1}^{N-1}\{(r_j-\lfloor \frac{i}{M}\rfloor)+(r_j-\lceil \frac{i}{M}\rceil)\}\\
\nonumber    =2r+3\sum\limits_{i=1}^{r-1}(r-i)\\
\nonumber    +2r^2(N-1)-\sum\limits_{r_j=1}^{r}\sum\limits_{i=1}^{N-1}(r_j-1)-\sum\limits_{r_j=1}^{r}\sum\limits_{i=1}^{N-1}r_j\\
\nonumber    +2r^2(M-1)-\sum\limits_{r_j=1}^{r}\sum\limits_{i=1}^{M-1}(r_j-1)-\sum\limits_{r_j=1}^{r}\sum\limits_{i=1}^{M-1}r_j\\
\nonumber    =2r+\frac{3}{2}[r(r-1)]+2r^2(M+N-2)\\
\nonumber    -\sum\limits_{r_j=1}^{r}[\sum\limits_{i=1}^{N-1}(2r_j-1)+\sum\limits_{i=1}^{M-1}(2r_j-1)]\\
\nonumber    =2r+\frac{3}{2}[r(r-1)]+2r^2(M+N-2)\\
\nonumber    -r(r+1)(M+N-2)+r(M+N-2)\\
=\frac{r^2}{2}(2M+2N-1)+\frac{r}{2}
\end{align}
Therefore, in order to efficiently utilize the available information, practical estimation of the second order statistics in the presence of sampling jitter requires the computation of the estimate using the self differences obtained by the individual samplers and its integration with the estimate obtained using the cross differences. 
\section{Computational Complexity}
\label{JITTER_multi_complexity}
In this section, complexity for the multiple period scenario is described along similar lines as described in~\cite{U_S_2} for the prototype co-prime scheme. Here, the cost for hardware implementation of autocorrelation estimation for the non-blind system in terms of the number of multiplications and additions is considered. Note that the computational complexity for the blind system is same as that of the ideal co-prime sampler with multiple periods described in~\cite{UVD_phdthesis}.

Let the number of multiplications required for the estimation of autocorrelation be denoted by $m_{b_r}(l)$ and $m_{nb_r}(l)$ for the blind and non-blind system respectively. Let the corresponding number of adders required be denoted by $a_{b_r}(l)$ and $a_{nb_r}(l)$. The subscript `$b_r$' refers to the blind system with $r$ co-prime periods, while `$nb_r$' refers to a non-blind system. The number of multiplications and additions are given in equation~\eqref{eq:multi_multipliers_l} and \eqref{eq:multi_adders_l} as a function of the difference value $l$ respectively. $z_{b_r}(l)$ and $z_{nb_r}(l)$ represent the number of contributors/ weight function after mapping $l\pm\frac{1}{2}\rightarrow l$ for the blind and the non-blind system with $r$ co-prime periods.
\begin{equation}\label{eq:multi_multipliers_l}
m_{b_r}(l)=z_{b_r}(l) ~\text{and}~ m_{nb_r}(l)=z_{nb_r}(l)
\end{equation}
\begin{align}\label{eq:multi_adders_l}
\nonumber   a_{b_r}(l)=z_{b_r}(l)-1;\{l|z_{b_r}(l)>0\}\\
a_{nb_r}(l)=z_{nb_r}(l)-1;\{l|z_{nb_r}(l)>0\}
\end{align}
Let $C_{M_{b_r}}$ and $C_{A_{b_r}}$ denote the total number of multiplications and additions respectively for a blind system. $C_{M_{b_r}}$ is the cumulative sum of $m_{b_r}(l)$ for $l\in[0,rMN-1]$, while $C_{A_{b_r}}$ is the cumulative sum of $a_{b_r}(l)$. It is same as that obtained for the ideal co-prime sampler with multiple periods derived in~\cite{UVD_phdthesis}.

Let $C_{M_{nb_r}}$ denote the total number of multiplications required for the non-blind system, and is the cumulative sum of $m_{nb_r}(l)$ for $l\in[0,rMN-1]$:
\begin{eqnarray}
\nonumber   &&C_{M_{nb_r}}\\
\nonumber       &=&\sum\limits_{l=0}^{rMN-1}m_{nb_r}(l)\\
\nonumber       &=&C_{M_{b_r}}+\frac{r^2}{2}(2M+2N-1)+\frac{r}{2}\\
\end{eqnarray}
Let $C_{A_{nb_r}}$ denote the total number of additions required for the non-blind system, and is the cumulative sum of $a_{nb_r}(l)$ for $l\in[0,rMN-1]$:
\begin{eqnarray}
\nonumber   C_{A_{nb_r}}&=&\sum\limits_{\{l|m_{nb_r}(l)>0\}}a_{nb_r}(l)\\
\nonumber       &=&C_{A_{b_r}}+\frac{r^2}{2}(2M+2N-1)+\frac{r}{2}\\
\end{eqnarray}
This is straightforward since the number of additional contributors available for estimation using the non-blind system is given by~\eqref{eq:jitter_multi_additional_contributors}. Hence, it justifies the above equations for $C_{M_{nb_r}}$ and $C_{A_{nb_r}}$.
\section{Conclusion}
This paper studies the difference set for the co-prime sampler with multiple periods in the presence of jitters, along similar lines as that of the prototype co-prime sampler which was previously studied. The number of distinct values in the sets are described. The number of contributors for autocorrelation estimation for the unmapped and mapped locations are provided. Finally the computational complexity is derived for autocorrelation estimation. The non-blind system has more number of contributors for estimation. 

In the future, researchers may investigate co-prime based schemes in the presence of jitters from a practical perspective. The focus of this paper was on the difference set. Other co-prime based structures can be studied along similar lines such as extended co-prime~\cite{4.62,UVD_Extended}, n-tuple co-prime~\cite{20.2} or multi-level prime arrays~\cite{20.3}, generalized co-prime arrays~\cite{4.44}, other multiple period structures~\cite{CAMPs}, etc.
%


%
%

\bibliographystyle{IEEEtran}   
\bibliography{refs}

\end{document}